\documentclass[journal]{IEEEtran}
\usepackage{cite}
\usepackage{graphicx}
\usepackage{epstopdf}
\usepackage{balance}
\usepackage{algorithm}
\usepackage{algpseudocode}
\usepackage{multicol}
\usepackage{multirow}
\usepackage{hyperref}
\usepackage[binary-units=true]{siunitx}

\ifCLASSINFOpdf

\else
\fi
\usepackage{amssymb,amsmath,amsthm}
\DeclareMathOperator*{\argmin}{arg\,min}
\DeclareMathOperator*{\argmax}{arg\,max}
\newtheorem{assumption}{Assumption}
\usepackage[x11names,dvipsnames,table]{xcolor} 
\usepackage{colortbl}
\usepackage{booktabs}
\usepackage{pbox}
\definecolor{def_blue}{cmyk}{0.541, 0.227, 0, 0.015}
\usepackage{todonotes}
\usepackage{tikz}
\usetikzlibrary{shapes,arrows,chains}
\usetikzlibrary{shapes.geometric, arrows}
\usetikzlibrary{shapes,snakes}
\usepackage[latin1]{inputenc}
\usetikzlibrary{calc,patterns}
\usetikzlibrary{quotes,angles}
\usetikzlibrary{shapes,arrows}
\usetikzlibrary{automata, positioning}
\usetikzlibrary{arrows.meta}
\tikzset{%
  >={Latex[width=2mm,length=2mm]},
            base/.style = {rectangle, rounded corners, draw=black,
                           minimum width=4cm, minimum height=1cm,
                           text centered, font=\sffamily},
  activityStarts/.style = {base, fill=blue!30},
       startstop/.style = {base, fill=red!30},
    activityRuns/.style = {base, fill=green!30},
         process/.style = {base, minimum width=2.5cm, fill=orange!15},
}

\tikzset{
block/.style={
  draw, 
  fill=blue!20, 
  rectangle, 
  minimum height=3em, 
  minimum width=2em
  },
  block1/.style={
  draw, 
  fill=blue!20, 
  rectangle, 
  minimum height=3em, 
  minimum width=2em
  },
sum/.style={
  draw, 
  fill=blue!20, 
  circle, 
  },
input/.style={coordinate},
output/.style={coordinate},
pinstyle/.style={
  pin edge={to-,thin,black}
  }
} 

\begin{document}
% Old title
% \title{Real-Time UAV Interaction: Challenging Missions Leveraging Autonomous Algorithm based on Automatically Generated Controllers}
%
\title{Real-time UAV Complex Missions Leveraging Self-Adaptive Controller with Elastic Structure}

\author{Mohamad~Abdul~Hady$^{1}$,~\IEEEmembership{Member,~IEEE,}
        Basaran~Bahadir~Kocer$^{2}$,~\IEEEmembership{Member,~IEEE,}
        Harikumar~Kandath$^{3}$,~\IEEEmembership{Member,~IEEE,}
        and~Mahardhika~Pratama$^{1}$,~\IEEEmembership{Senior~Member,~IEEE.}% <-this % stops a space
\thanks{$^{1}$School of Computer Science and Engineering, Nanyang Technological University, 50 Nanyang Avenue, Singapore. Email: mpratama@ntu.edu.sg.}%         
\thanks{$^{2}$Aerial  Robotics  Laboratory,  Imperial  College  London,  South  Kensington, London. Email: b.kocer@imperial.ac.uk.}%
\thanks{$^{3}$ International Institute of Information Technology, Hyderabad-500032, India. E-mail: harikumar.k@iiit.ac.in.}
}


\markboth{Under Review in IEEE Transactions}%IEEE Transactions  on Systems, Man, and Cybernetics: Systems}%
{Shell \MakeLowercase{\textit{et al.}}: Bare Demo of IEEEtran.cls for IEEE Journals}



\onecolumn

\maketitle

\textit{  $\copyright$ 2020 IEEE.  Personal use of this material is permitted.  Permission from IEEE must be obtained for all other uses, in any current or future media, including reprinting/republishing this material for advertising or promotional purposes, creating new collective works, for resale or redistribution to servers or lists, or reuse of any copyrighted component of this work in other works.}\\
\clearpage
\twocolumn
\begin{abstract}
The expectation of unmanned air vehicles (UAVs) pushes the operation environment to narrow spaces, where the systems may fly very close to an object and perform an interaction. This phase brings the variation in UAV dynamics: thrust and drag coefficient of the propellers might change under different proximity. At the same time, UAVs may need to operate under external disturbances (e.g., wind gusts) to follow time-based trajectories. Under these challenging conditions, a standard controller approach may not handle all missions with a fixed structure, where there may be a need to adjust its parameters for each different case. With these motivations, practical implementation and evaluation of an autonomous controller applied to a quadrotor  UAV are proposed in this work. A self-adaptive sliding mode controller (SMC) based on a composite control scheme, where a combination of a sliding surface with structure of a PID controller and evolving neuro-fuzzy control is used. The parameter vector of the neuro-fuzzy controller is updated adaptively based on the sliding surface of the SMC. The autonomous controller possesses a new elastic structure, where the number of fuzzy rules is self-organized based on bias and variance balance.  The interaction of the UAV is experimentally evaluated in real time considering the ground effect, ceiling effect and flight through a strong fan-generated wind while following time-based trajectories.
\end{abstract}

\begin{IEEEkeywords}
Composite controller, evolving architecture,  sliding mode controller, trajectory tracking, unmanned aerial vehicle interaction.
\end{IEEEkeywords}

% \fi
\IEEEpeerreviewmaketitle

\section{Introduction}

The use of UAVs has already explored within the flying camera concept, where the system flies in the air in a passive manner [1]. Recently, the expectation of UAVs has shifted towards interaction operations in which the flying robot may interact and collaborate with cooperative or uncooperative objects, other robots and people in order to efficiently plan and execute the tasks. In most of these cases, UAVs need to conduct the flight in close proximities to objects, ground, ceiling or people [2][3].

When the flying robots operate near objects, their aerodynamic properties might significantly change. For an autonomous aerial robot to anticipate these changes, it is required to include responsive, adaptive and resilient techniques. In this context, there is a need to inspire from nature where flying animals have attributes, including self-orientation and survival by predicting potential dangers. Since the aerodynamic properties of the rotors significantly change, the performance of the aerial robot may degrade in close proximity, which might result in a crash if the aforementioned properties are not included in the control system.

{One of the most common close proximity flight conditions is associated with the ground effect which can degrade the system performance during landing and proximity flight to the ground. A model based approach that uses  a precise mathematical model of the ground disturbance effect obtained through experiments is provided in [4]. But the developed disturbance model is specific to the propeller specifications and the configuration of the UAV. However, there is still a safety margin, where the ground effect creates an additional lift on the system { in the open air}. On the other hand, {its counterpart -} ceiling effect pulls the flying object to the ceiling. Without a proper control strategy, this phenomenon can lead to a crash. In this context, the focus had been directed towards generating an alternative path when the ceiling effect is detected similar to [5]. An exact opposite approach is explored recently where the UAV is driven to be in contact with the ceiling [6]-[8]. However, there is still a need to examine close proximity flight by compensating the variable aerodynamic interactions. Since the aerodynamic coefficients of the propellers are changing during the flight in close proximity to the ceiling, a variable control allocation matrix is proposed in [9].} {As compared to the case where the fixed control allocation matrix is used, the performance is increased. However, mapping and measuring these values are more troublesome in real-time experimental evaluations. A recent model-based approach is proposed in [10], where the model predictive controller (MPC) is utilized based on a nominal model of the system. The presented approach is promising, but it requires a precise model and the optimization brings an additional computational cost [11]. Estimation based techniques employ a disturbance observer to estimate the disturbances acting on the system. Disturbance observer increases the computational load and also degrade transient performance  due to the lag in the estimated disturbance and true disturbance [12]. } 

A potential interaction can arise during the flight close to a vertical wall. {However, it is not a dominant effect as compared to the flights where the additional aerodynamic forces are effective on the thrust direction.} A more challenging task, e.g., flying in a UAV flock in tightly close formations, can be represented by flying through a fan along the horizontal axis. In this context, there are some alternatives already available in the literature to suppress the disturbance while tracking the commanded trajectories including MPC [13][14], dynamic inversion control [15][16], and adaptive control [17][18]. Among them, SMC is one of the strong candidates to handle internal and/or external disturbances and the system uncertainties [19][20]. In this context, SMC-based approaches are investigated in a simulated environment for the attitude and the position loops of the UAVs [21]. As a real-time experiment, a finite time stabilizing SMC is applied to a ground-based experimental setup [22]. However, the evaluation is restricted to the attitude channel of the system. Recently, a combination of PID and super twisting SMC are utilized to the UAV for the position control against disturbances [23]. Without loss of generality, the aforementioned nonlinear control approaches are associated with the accuracy of the plant dynamics and information about the operating environment. 

When the system goes into an interaction with the ceiling or the ground, the model of the UAV changes (e.g., variable thrust characteristics of the propeller). Similarly, flying through a fan might significantly affect the system. {Several works have been addressed wind disturbance using a fan. A nonlinear adaptive state feedback controller is designed using Lyapunov-based backstepping techniques to maintain the path following to suppress wind disturbance [24]. Besides, a disturbance observer is evaluated to compensate for the wind disturbance generated by a fan in [25] and [26]. However, the aforementioned approaches are designed for a particular task and required a precise system model. Composite adaptive controllers are shown to reject disturbances that are large in magnitude and bandwidth [27]. But the composite adaptive controller mentioned in the literature for UAV control requires manual tuning of a set of parameters depending upon the control objective to be achieved. \\
Instead of the model-based controller, there are some learning-based approaches considered for UAV applications including fuzzy logic systems [28], neural networks [29][30], and neuro-fuzzy-based control approaches [31][32]. Furthermore, there are some approaches by combining different approaches, e.g., MPC with fuzzy logic control [33]. However, available learning-based approaches mostly using fixed structured networks with a fixed number of rules {that} do not provide a flexible learning-scheme [34]. Such controllers often lead to poor performance while performing complex flight missions without adequate retraining of the network or by using different combinations of the controller gains.}

 {Recently, some evolving neuro-fuzzy controllers are proposed to tackle the varying system dynamics [35]. According to numerical investigations, it outperforms the PID (linear) and TS Fuzzy (nonlinear with fixed structure) controller. Nevertheless, it is evaluated within a simulation environment and it includes numerous {parameters} to be adjusted, which is arduous to be implemented in a real-time experiment with limited computational power. Another evolving controller was proposed based on a neuro-fuzzy structure, which is a combination of an autonomous learning controller with the SMC  [36]. This controller  not only achieved better performance than the fixed structure ones but also designed to be significantly less number of parameters by using hyperplane based membership function.} {However, this work is evaluated only in a simulation environment for altitude and attitude tracking of a multirotor. Motivated by having a more flexible learning scheme, our contribution is focused on an evolving neural network with a combination of fuzzy and sliding mode control. It does not require information about the changing internal and external parameters. Different from the precisely identified model-based control approaches or data-hungry and fixed structured neural network-based controllers,} the proposed learning scheme is based on a three-layered neuro-fuzzy system with autonomous rule generation feature. By estimating the bias and variance of the network, the structure is self-organized based on the trade-off between under-fitting and over-fitting conditions. 
 
In this study,  an autonomous controller with an elastic structure is proposed to solve the aforementioned problems inherent to UAV interaction and other challenging missions. {Unlike in [36], where the input to the neuro-fuzzy structure contains the derivative of the error, here integral of error is used yielding better steady-state performance. In addition, a stability analysis of the composite controller is provided based on first order system dynamics employing velocity control.} In summary, the following novelties are presented in this study:

\begin{itemize}
    \item For the first time, a composite controller comprising of a sliding surface with the structure of a PID controller and an evolving neuro-fuzzy network structure is applied to an aerial robot performing complex missions.
    \item By estimating the network bias and variance, a new elastic structure based on a neuro-fuzzy controller is implemented and evaluated for an aerial robot experimentally in real-time.
    \item The performance of the proposed controller is evaluated under the ground and ceiling effect as well as circle and 8-shaped time-based trajectory tracking under wind disturbance. 
    \item  {The proposed controller has been  validated under the outdoor environment with wind disturbances using realistic simulator namely Ardupilot-SITL (software in the loop simulator) using Dronekit API in which the  simulation codes\footnote{Simulation codes are available in the following link: https://github.com/ContinualAL/PAC-with-Dronekit-SITL} and flight videos\footnote{Experimental video is available in:\url{https://youtu.be/-nGhMzvB7oE}} are provided in the external link.}
    \item {The proposed method is able to work in different mission scenario without further tuning.}
\end{itemize}

This paper is organized as follows: Section \ref{sec_problem} formally describes the problem of trajectory tracking in the presence of internal and external disturbances. Section \ref{sec_design} introduces the detailed structure, mechanism, and update of the autonomous controller, namely Parsimonious Autonomous Controller (PAC). In Section \ref{sec_simulations}, the numerical simulation is investigated and discussed. The experimental results and the corresponding analysis are presented in Section \ref{sec_experiments}. The discussion of the performance evaluation summary is provided in  Section \ref{sec_discussion}. And the last part, Section \ref{sec_conclusion}, concludes the paper, including the future work.

\section{Problem Formulation}\label{sec_problem}
\begin{figure}[!t]
    \centering
  \begin{tikzpicture}[auto,>=latex']
    % We start by placing the blocks
    \node [input, name=input] {};
    \node [sum, right = of input] (sum1) {};
    \node [block, right = of sum1] (pac) {$\kappa(.)$};
    %\node [sum, right = of pac] (sum2) {};
    %\node [block, right = of sum2] (controller) {$PID$};
    \node [block, right = of pac] (system) {UAV};
    \node [sum, right = of system] (sum3) {};
    
    % We draw an edge between the controller and system block to 
    % calculate the coordinate u. We need it to place the measurement block. 
    %\draw [->] (controller) -- node[name=u] {$u_c$} (system);
   % \node [input, above =of sum2](d1) {};
     \node [input, above =of sum3](d2) {};
   % \node [output, right =of system] (output) {};
     \node [block, right = of sum3
            ] (system1) {$\frac{1}{s}$};
    \node [output, right =of system1] (output1) {};
    % \node [, below =of controller] (a1) {$y_1(k)$};

    % Once the nodes are placed, connecting them is easy. 
    \draw [draw,->] (input) -- node {$\eta_r(t)$} (sum1);
    %\draw [draw,->] (d1) -- node {$\dot{\eta}_{d}(t)$} (sum2);
   \draw [draw,->] (d2) -- node {$\dot{\eta}_{d}(t)$} (sum3);
    \draw [draw,->] (sum1) -- node {$e_{\eta}(t)$} (pac);
    %\draw [draw,->] (pac) -- node {$\dot{\eta}_r(t)$} (sum2);
    \draw [->] (pac) -- node {$\dot{\eta}_r(t)$} (system);
   % \draw [->] (system) -- node [name=y] {$\dot{\eta}(t)$}(output);
    \draw [->] (system) -- node {$\dot{\eta}_o(t)$} (sum3);
   \draw [->] (sum3) -- node {$\dot{\eta}(t)$} (system1);
    \draw [->] (system1) -- node [name=y1] {$\eta(t)$}(output1);
    %\draw [->] (y) |- (sum2);
    %\draw [->] (a1) -- (controller);
    \draw [->] (y1) -- ++(0,-3cm) -| node[pos=0.99] {$-$} 
        node [near end] {} (sum1);
        %  \draw [->] (y) -- ++(0,-2cm) -| node[pos=0.99] {$-$} 
        % node [near end] {$r$} (sum2);
\end{tikzpicture}
    \caption{First order control architecture for the UAV with disturbances.}
    \label{mav1con2}
\end{figure}
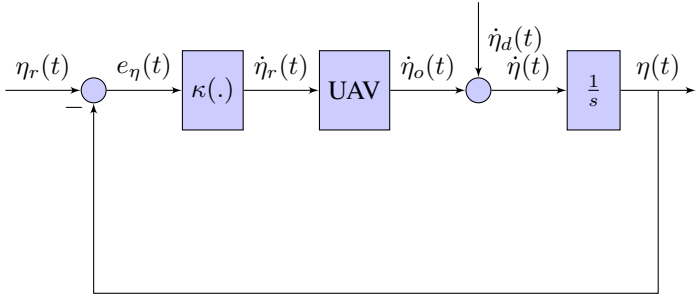

Based on the mission requirements, the UAV needs to navigate in challenging environments like in the proximity of a wall, ceiling or ground, or under external wind disturbances. In a region having obstacles, following a straight line trajectory connecting the current position and the desired position is not always feasible. Navigating in such regions requires the following of a circular or an 8-shaped trajectory [37]. Consider the first order dynamics of the UAV for trajectory control as given below.
\begin{equation}
    \dot{\eta}(t)=\dot{\eta}_o(t)+\dot{\eta}_{d}(t)
    \label{eqd1}
\end{equation}
where ${\eta}(t)$ is the position of the UAV, $\dot{\eta}_{o}(t)$ is the UAV velocity (without disturbance) and  $\dot{\eta}_{d}(t)$ is the disturbance velocity input as shown in Fig. \ref{mav1con2}. Here the assumption is that the UAV is enabled with an inner loop velocity control to track the reference velocity input, i.e. $\dot{\eta}_{o}(t)\approx\dot{\eta}_{r}(t)$. The unknown disturbance velocity input is due the effect of the wind disturbances and interaction effects. The controller $\kappa(.)$ is a mapping that relates the tracking error $e_{\eta}(t)$ to the reference velocity input $\dot{\eta}_{r}(t)$ as represented below.
\begin{equation}
 \dot{\eta}_{r}(t)=\kappa\big(e_{\eta}(t)\big).
\label{eqd2}
\end{equation}
The design of the controller possesses some challenges as stated below.
\begin{itemize}
    \item The reference trajectory $\eta_r(t)$ is time-varying and having a nonlinear time-dependent function. Moreover, the use of a controller with a fixed structure might not provide efficiency for tracking different reference trajectories under disturbances in an unknown environment [38].
    \item The power spectra of the wind disturbances are quite different from that of interaction effects. The controller should handle disturbances with unknown power spectra and those belonging to different categories.
    \item {The lift and pull forces produced by the ground and ceiling effects are not measured. This disturbance type is categorized as an unknown disturbance.}
  \end{itemize}

The use of nonlinear robust control techniques like SMC requires information about the upper bound of the disturbance acting on the system [39]. Using a conservative upper bound for different kinds of disturbances might result in the usage of large control effort. For tracking time varying trajectories in the presence of wind disturbances and interaction effects, an autonomous navigation algorithm with an evolving structure is proposed in this paper. The proposed trajectory tracking controller is composite, consisting of a fixed sliding mode controller and a neuro-fuzzy system with varying number of fuzzy rules. The equations of motion for the trajectory tracking controller and the underlying assumptions are described below.
 
%\subsection{Equations of Motion}
The position of the UAV in the inertial frame is denoted by $\eta(t)$=($x,\,y,\,z$). The expanded version of (\ref{eqd1}) - first order dynamics relating the reference velocity inputs $\dot{\eta}_r(t)$=($v_{xu},\,v_{yu},\,v_{zu}$), disturbance velocity inputs $\dot{\eta}_d(t)$=($v_{xd},\,v_{yd},\,v_{zd}$) and the inertial position is given in (\ref{teq11}), (\ref{teq12}), and (\ref{teq13}):
\begin{align}
     \dot{x}=v_{xu}+v_{xd}, \label{teq11} \\   
     \dot{y}=v_{yu}+v_{yd}, \label{teq12} \\  
     \dot{z}=v_{zu}+v_{zd}. \label{teq13}   
\end{align}

The objective  is to develop an autonomous control algorithm to generate the reference velocity inputs 
($v_{xu},\,v_{yu},\,v_{zu}$), to track the desired trajectory $\big(x_r(t),\,y_r(t)),\,z_r(t) \big)$  in the presence of unknown  disturbance velocity inputs denoted by ($v_{xd},\,v_{yd},\,v_{zd}$).  The  reference velocity inputs are tracked by the conventional inner loop PID controllers, where a cascaded structure considering time scale differences between the translation and attitude dynamics is considered similar to [40]. The error vector and the error dynamics are given in (\ref{phm1}) and (\ref{phm2}),
\begin{equation}
\footnotesize
\Big[e_x(t),\,e_y(t),\,e_z(t) \Big]^{\rm T} = \Big[x_r(t)-x(t),\,y_r(t)-y(t),\,z_r(t)-z(t) \Big]^{\rm T},
     \label{phm1}
\end{equation}
\begin{equation}
 \left (\begin {array}{c} \dot{e}_x(t)\\  \dot{e}_y(t) \\ \dot{e}_z(t) \end {array} \right)=  \left (\begin {array}{c} \dot{x}_r(t)-v_{xu}-v_{xd}\\  \dot{y}_r(t)-v_{yu}-v_{yd} \\\dot{z}_r(t)-v_{zu}-v_{zd} \end {array} \right).
 \label{phm2}
\end{equation}
The following assumptions are utilized in this paper. 
\begin{assumption}\label{ass_traj}
The reference trajectory $\big(x_r(t),\,y_r(t),\,z_r(t) \big)$ is uniformly continuous. 
\end{assumption}
\begin{assumption}\label{ass_bound}
The disturbance velocity inputs indicated by ($v_{xd},\,v_{yd},\,v_{zd}$) are bounded with unknown power spectra. \end{assumption}
\begin{assumption}\label{ass_stab}
{The UAV is equipped with inner loop controller that can track the  velocity command generated by the proposed controller.} \end{assumption}
These Assumptions enable the UAV to track the desired trajectory using bounded reference velocity inputs generated by the proposed controller.

\section{Parsimonious Autonomous Controller (PAC)}\label{sec_design}
\begin{figure}[!t]
\centerline{\includegraphics[width=8.5cm, height=4.5cm]{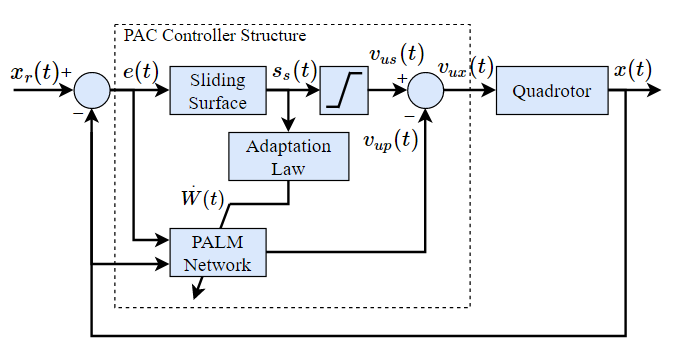}}
\caption{UAV control system block diagram for $x$-axis with PAC.}
\label{figblock}
\end{figure}

\begin{figure}[b]
\centerline{\includegraphics[width=9cm, height=6.4cm]{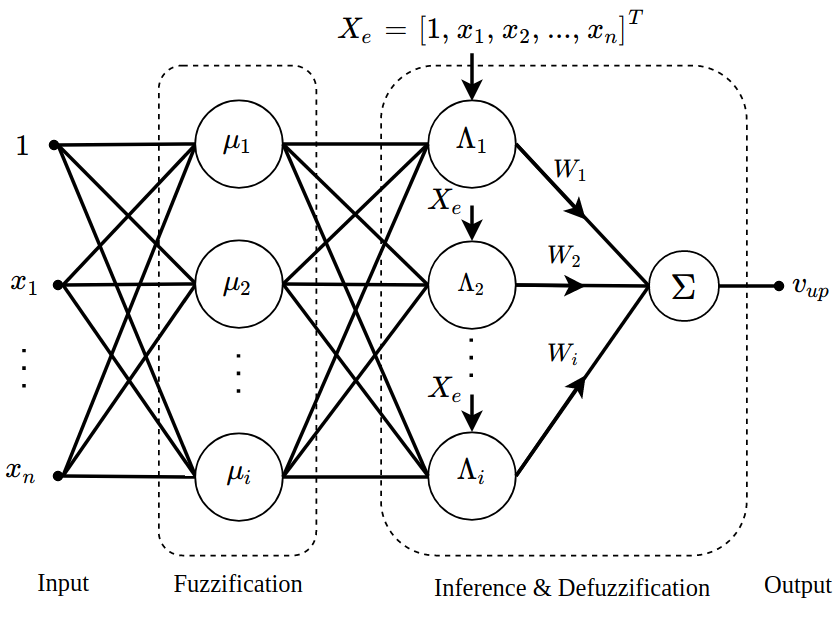}}
\caption{Neuro-fuzzy structure of PALM.}
\label{palmstruct}
\end{figure}

The structure of a quadrotor UAV control scheme using PAC is given in Fig. \ref{figblock} for the case of $x$ coordinate of the trajectory \footnote{The pseudocode for the PAC is provided in the supplementary material.}. The same architecture is used for the case of $y$ and $z$ coordinates and hence not discussed here separately.
% Let $p_s(t)$
% denote the system output, $p_r(t)$ denote the reference input to
% be tracked and $e(t)=p_r(t)-p_s(t)$ be the tracking error. For
% the case of trajectory tracking $p_s(t) \in \{x(t),\,y(t),\,z(t)\}$, $p_r(t) \in \{x_r(t),\,y_r(t),\,z_r(t)\}$ and $e(t) \in \{e_x(t),\,e_y(t),\,e_z(t)\}$.
%
PAC is constructed from SMC with a modification in the equivalent control input part to be adaptable. The sliding surface \big($s_s(t)$\big) possesses the structure of a PID controller as given below
\begin{equation}\label{sl_surf}
s_s(t)=k_1 e_x(t) + k_2 \int_{0}^{t} e_x(\tau) d\tau + k_3 \dot{e}_x(t).
\end{equation}
where the gains $k_1>0$, $k_2 > 0$ and $k_3 \geq 0$. The velocity input from the sliding surface
\begin{equation} \label{smc}
    v_{us}=\text{sat}(s_s(t)).
\end{equation}
The saturation operation  for any function $f(t)$ with $a_m>0$ is defined as given below
\begin{equation}
\text{sat}(f(t))=\left\{
                \begin{array}{ll}
                  f(t), \,\, \textrm{if} (|f(t)| \leq a_m)\\
                  a_m \text{sign}(f(t)), \,\, \textrm{if} (|f(t)|> a_m).
                \end{array}
              \right.
\label{scon2}
\end{equation}
The adaptive structure part of the PAC is inherited from autonomous deep learning (ADL) [41] to generate the input signal considering an unknown disturbance acting on the system. The network architecture of PAC is adapted from PALM [42] given in  Fig. \ref{palmstruct}. In this structure, the number of membership functions is equal to the number of rules denoted by $r(t)$. Unlike the conventional fixed structure of neuro-fuzzy, where the number of rules is pre-defined, here $r(t)$ is the number of rules at a given time $t$ and is time-varying due to the dynamics changing. The neuro-fuzzy structure of PALM consists of three layers, namely: (i) input layer, (ii) fuzzification, (iii) inference and defuzzification.  Each of these layers is discussed separately below.

\textit{(i) Input layer}: The input is a combination of extended input (equal to 1), error, integral error, and the system output $x(t)$. The resulting input vector is given below

\begin{equation}
 X_{e}=[1,e_x(t),\int_{0}^{t} e_x(\tau) d\tau,x(t)]^{\rm T}.
\end{equation}

 \textit{(ii) Fuzzification}: To transform the crisp value inputs into a fuzzy value, a hyperplane-based membership function $\mu_{i}$ is selected in the fuzzification layer as given in (\ref{miu}) with $i \in \{1,\,2,\,3,\,..,\,r(t) \}$
 
 \begin{equation}
\mu_{i}=\exp\left(-\alpha \frac{\delta_{i}}{\max(\delta_{i})}\right).
\label{miu}
\end{equation}

 \noindent The parameter $\alpha >0$ and $\delta_i$ is the normalized distance of the $x_r$ from the $i$-th hyperplane ($W_{i}X_{e}=0$) as shown below
 
 \begin{equation}
\delta_i=\left\lvert\frac{x_{r}-(W_{i}X_{e})}{\sqrt{W_{i}W_i^{\rm T}}}\right\rvert.
\label{dist}
\end{equation}

\textit{(iii) Inference and Defuzzification} : Takagi-Sugeno fuzzy inference system is applied in this controller with each rule (Rule-$i$) statement as the following statement

\begin{equation}
\resizebox{.85\hsize}{!}{$
\textrm{Rule-}i\;:\:\mathtt{\textrm{IF\:}}X_{e}\:\textrm{is Close to\:}\mu_{i}\:\textrm{ THEN }\:O_{i}=W_{i}\Lambda_i$}.
\label{rule}
\end{equation}

The output of each Rule-$i$ ($O_i$) is obtained by multiplying the network parameter ($W_i$) with $\Lambda_{i}$ calculated from (\ref{lambdai}). It is observed that the proposed algorithm incurs parsimonious network parameters compared to the conventional neuro-fuzzy controller in the literature because it is free of rule premise parameters. The hyperplane functions are used in both the rule premise and the consequent part.

\begin{equation}
\Lambda_i=\lambda_i X_{e},
\label{lambdai}
\end{equation}

\noindent where the $\lambda_i$ is the normalized firing strength of rule-$i$ calculated as:
\begin{equation}
    \lambda_i=\frac{\mu_{i}}{\sum_{i=1}^{R}(\mu_{i})}.
\end{equation}
The final defuzzified (crisp) output is given by
\begin{equation}\label{defuz}
v_{up}=\sum_{i=1}^{R}(O_{i}).
\end{equation}
Furthermore, the reference velocity input to be given to the system ($v_{ux}$) shown below is the difference of the output of sliding surface given in (\ref{smc}) and the defuzzified output given in (\ref{defuz})

\begin{equation}\label{eq:controlsignal}
v_{ux}=v_{us}-v_{up}.
\end{equation}

The adaptation of PAC is implemented into two parts discussed in the next two sub-sections. The first part is the adaptation of the elastic network structure (adjusting the number of rules $\big(r(t)\big)$ regulated by the growing and pruning mechanism. The second part is the adaptation of network parameters $\big(W_i(t)\big)$ updated to minimize the cost function derived from the sliding surface output.

\subsection{Elastic network structure adaptation mechanism}

PAC starts its learning process from scratch with the absence of a predefined network structure. Its structure is self-evolving with respect to the generalization power, referring to the bias and variance trade-off in the machine learning literature. The expression of the mean squared error (MSE) for the network prediction is given below.
\begin{equation}\label{eqn:MSE}
\text{MSE}=\sum_{t^*=0}^T\frac{1}{T}(x[t^*]-x_p[t^*])^{2},
\end{equation}
where $x[t^*]$ and $x_p[t^*]$ is the sampled system output (UAV position along the x-axis) and the prediction output from the network respectively, within sampling interval $0<t^* \le T$. This equation raises further issues in the single-pass learning scenario due to: (i) the requirement of memory to record all sampled data points to represent the network capacity, (ii) even though it can be calculated recursively without accumulating all of the sampled data, it can not represent the network capability for unseen input data samples, (iii) minimization of (\ref{eqn:MSE}) leads to an unbiased estimator but risks on the overfitting situation. Thus, the network significance is re-formulated as given below
\begin{equation}\label{eqn:NS1}
N_s=\int_{-\infty}^{\infty} \big(x-x_p(X_e)\big)^2f_p(X_e)d X_e,
\end{equation}
where $f_p(X_e)$ is the probability density function of the network input. The equation (\ref{eqn:NS1}) is modified as shown below where $E[.]$ is the expectation operator
\begin{equation}\label{eqn:NS2}
    N_s =E[(x-x_p)^{2}].
\end{equation}

After implementing some algebraic operations, the following equation is obtained
\begin{equation}\label{eqn:NS3}
    N_s =(x-E[x_p])^{2}+(E[x_p^2]-E[x_p]^2).
\end{equation}

The first term of (\ref{eqn:NS3}) denotes the bias  ($\beta_{x}^2$) and the second term corresponds to the variance ($\vartheta_x$) of the network output as written below.
\begin{equation}\label{eqn:final_NS}
    N_s =\beta_{x}^2+\vartheta_x.
\end{equation}

To establish the bias and variance calculation in PAC, a unity rule firing strength ($\lambda_i$) is assumed. It can simplify the network output as $x_p=\sum_{i=1}^{r(t)} W_i X_e$, because the expectation of the normalized firing strength does not have unique solution as the nature of hyperplane-based membership function. The expectation of $x_p$ is re-written as given below.
\begin{equation}
E[x_p]=\int_{-\infty}^{\infty} X_e W_i\, f_p(X_e)\, dX_e.
\end{equation}
Assuming that the probability density function of $X_e$ is normally distributed, the above equation is modified as given below
\begin{equation}\label{eqn:expectx}
       E[x_p]=W_i\mu_{x}.
\end{equation}
where $\mu_x$ is the mean of extended input $X_e$, that is originating from the integration of $X_e$ overtime $-\infty <t<\infty$. Note that $\mu_{x}$ in (\ref{eqn:expectx}) is recursively updated and also adapts to changing learning environments. The use of a mixture of Gaussians can be implemented to address the dependence on the normal distribution but costs high computational complexity and not suitable for real-time implementation. 

By referring to  equation (\ref{eqn:NS3}), the variance is obtained from $\vartheta_x=E[x_{p}^2]-E[x_p]^2$, where the second term is simply $E[x_p]^2=E[x_p]\times E[x_p]$ and the first term can be formulated as $E[x_{p}^2]=W_i(\mu_x)^2$. Combining these results, the final expression for $N_s$ formula is obtained. Since  $N_s$ depends on both bias and variance, a high value of $N_s$ indicates that either the high bias (over-simplified network with measly rules) or the high variance (over-complex network with numerous rules). To maintain the network capacity with respect to the complexity, the concept of bias-variance trade-off is applied to this structural evolution mechanism. The network structure has a low approximation capacity if the $\beta_x$ value is high. Therefore, the number of rules has to increase. On the contrary, a high value of $\vartheta_x$ is due to the excess number of rules. In this case, the number of rules has to reduce. Both of these mechanisms are applied to adapt the network elastically in order to reach the optimum approximation capacity. Two criteria for rule growing and pruning are formulated as:

\textit{(i) Rule growing:} The criteria for rule growing is given below.
\begin{equation}\label{eq:add_R}
\mu_{\beta}(t)+\sigma_{\beta}(t)\ge \mu_{\beta}^{\min}+\varGamma \sigma_{\beta}^{\min},
\end{equation}
where $\mu_{\beta}(t)$ is the mean of bias and $\sigma_{\beta}(t)$ is the standard deviation of the system output at a time $t$. On the right side, the notation $\mu_{\beta}^{\min}$ is the minimum value of the mean bias and the $\sigma_{\beta}^{\min}$ is the minimum standard deviation of bias which have ever existed. This minimum value is reset each time the growing condition is satisfied.  

\textit{(ii) Rule pruning:} Similar notations are used to express the rule pruning condition as shown below. 
\begin{equation}\label{eq:prune_R}
\mu_{\vartheta}(t)+\sigma_{\vartheta}(t)\ge \mu_{\vartheta}^{\min}+\zeta \sigma_{\vartheta}^{\min}.
\end{equation}
The minimum value of the mean and standard deviation of variance is reset whenever rule pruning is executed. The growing and pruning parameter is designed to be adaptable as $\varGamma=1.5\exp(-\beta_e^2)+0.5$ and $\zeta=1.5\exp(-\vartheta_x)+0.5$. These parameters vary between 0.5 and 2 and behave as the confidence interval ranges from 0.5$\sigma$($\approx38\%$) to 2$\sigma (\approx95\%)$. If the network bias or variance is high - the confidence level is lower, conversely, if these are low - the confidence level is higher. That is, a new rule is added in the case of high bias whereas an inactive rule is removed in the case of high variance.   

The rule pruning condition is based on the rule significance parameter ($RS_i$) formulated from the absolute value of the network output expectation in (\ref{eqn:expectx}). The $RS_i$ for $i$-th rule is calculated as follows:
\begin{equation}\label{rulesig}
    RS_i=\left\lvert\vert W_i \mu_x \right\rvert\rvert_1 .
\end{equation}
The rule of interest is based on having the lowest statistical contribution (\ref{rulesig}). Hence, the $j$-th rule is pruned if the condition in (\ref{rulemin})is satisfied
\begin{equation}\label{rulemin}
    \argmin_{i\in[1,2,3,..,r(t)]}(RS_i) = RS_{j}.
\end{equation}

This elastic learning feature is capable of tracking any variations of plant dynamic without redesigning from scratch - common practice for a fixed and offline controller. In other words, it copes with different flight missions autonomously without applying pre-fixed controller gains.

\subsection{Adaptation of network parameters}

The hyper-plane parameters corresponding to the winning rule is only updated at each sampling instant. The winning rule is evaluated with the same network significance as outlined in (\ref{rulesig}). The $i$-th rule is the winning rule if the condition below is satisfied.
\begin{equation}
    \argmax_{k\in[1,2,3,..,r(t)]}(RS_{k}) = RS_{i}.
\end{equation}

The adaptation of the winning rule parameters at a  time instant $t$ $\big(W_i(t)\big)$ is designed to bring the tracking error to the sliding surface. Therefore, the cost function is chosen from the sliding surface function and it is assumed as $v_{us}(t)\approx s_s(t)$. By recalling equations (\ref{rule}), (\ref{defuz}), and (\ref{eq:controlsignal}), the cost function is described below.
 \begin{equation}
 \begin{split}
     J(t) & =\int_{0}^{t}{s_{s}(\tau)^2\,d\tau}, \\
     & = \int_{0}^{t}{(v_{ux}(\tau)+v_{up}(\tau))^2\,d\tau}, \\
     & = \int_{0}^{t}{(v_{ux}(\tau)+W_i(\tau)\Lambda_i(\tau) )^2\,d\tau}. \\
 \end{split}
\end{equation}
The condition for optimality can be stated as $\frac{\partial J(t)}{\partial W_i(t)}= 0$. The adaptation rule given in (\ref{dj_dw}) is obtained by following the derivation  given in [43]
\begin{equation}\label{dj_dw}
 \begin{split}
     \frac{\partial J(t)}{\partial W_i(t)} & = 0, \\
     0 & = \int_{0}^{t}{\Lambda_i(\tau)v_{ux}(\tau)\,d\tau}+\int_{0}^{t}{\Lambda_i(\tau)\Lambda_i(\tau)^{\rm T}\,d\tau\,W_i(t)}, \\
     W_i(t) & = -\bigg[\int_{0}^{t}{\Lambda_i(\tau)\Lambda_i(\tau)^{\rm T}\,d\tau}\bigg]^{-1}\int_{0}^{t}{\Lambda_i(\tau)v_{ux}(\tau)\,d\tau}. \\
\end{split}
\end{equation}
Let $P_i(t)=\big[\int_{0}^{t}{\Lambda_i(\tau)\Lambda_i(\tau)^{\rm T}\,d\tau}\big]^{-1}$, then (\ref{dj_dw}) can be re-written as:
 \begin{equation}\label{eq_w}
 \begin{split}
     W_i(t) & = -P(t)\,\int_{0}^{t}{\Lambda_i(\tau)v_{ux}(\tau)\,d\tau}. \\
 \end{split}
\end{equation}
By taking the derivative of the above equation and simplifying the mathematical expression gives the following formula for weight update
 \begin{equation}\label{eq:update_W}
 \dot{W_i}(t)=- P_i(t)\Lambda_i(t)s_s(t),\;\;\text{where }\;W(0)=W_{0}\in\Re^{R\times n}.
 \end{equation}
The matrix $P_i(t)$ is updated recursively without recording all data samples as given below
 \begin{equation}\label{eq:update_P}
 \dot{P_i}(t)=-P_i(t)\Lambda_i(t)\Lambda_i^{\rm T}(t)P_i(t),
 \end{equation}
where $P_i(0)=P_{i_0}>0\,\in\Re^{n\times n}$.

\subsection{Stability analysis}

The stability analysis for the closed loop first order dynamics given in (\ref{teq11}) is presented in this section. Let
\begin{equation}
    x_1=\int_{0}^{t} e_x(\tau) d\tau
\end{equation}
Then
\begin{equation}
    \dot{x}_1=x_2=e_x(t)
\end{equation}
Using the dynamics in (\ref{teq11}),
\begin{equation}
   \dot{e}_x(t)= \dot{x}_r(t)-v_{xu}-v_{xd}
\end{equation}
where
\begin{equation}
    v_{xu}=v_{us}-v_{up}
\end{equation}
and the control input corresponding to the sliding surface,
\begin{equation}
    v_{us}=k_1 x_2 + k_2 x_1 + k_3 \dot{x}_2
    \label{ss1}
\end{equation}
the following state space model can be obtained,
\begin{equation}
\left (\begin {array}{c} \dot{x}_1\\  \dot{x}_2 \end {array} \right)=\left (\begin {array}{cc} 0&1\\  \frac{-k_2}{1+k_3}&\frac{-k_1}{1+k_3} \end {array} \right)\left (\begin {array}{c} x_{1}\\  x_{2} \end {array} \right)+\left (\begin {array}{c} 0\\  \frac{1}{1+k_3} \end {array} \right)U
    \label{p1a}
\end{equation}
where $U=\dot{x}_r(t)-v_{xd}+v_{up}$.
From Assumptions \ref{ass_traj} and \ref{ass_bound}, the term $\dot{x}_r(t)-v_{xd}$ is bounded.
So the system dynamics given in (\ref{p1a}) is stable if the matrix
$A_{xe}$ given by
\begin{equation}
 A_{xe}=\left (\begin {array}{cc} 0&1\\  \frac{-k_2}{1+k_3}&\frac{-k_1}{1+k_3} \end {array} \right)
    \label{p2a}   
\end{equation}
is Hurwitz and the output from PAC denoted by $v_{up}$ is bounded. The matrix $A_{xe}$ is Hurwitz for $k_1>0$, $k_2>0$ and $k_3\geq0$. 
The weight update rule for PAC is obtained by minimizing the following performance index.
\begin{equation}
 \begin{split}
     J(t) & =\int_{0}^{t}{s_{s}(\tau)^2\,d\tau}, \\
     & = \int_{0}^{t}{(v_{ux}(\tau)+v_{up}(\tau))^2\,d\tau}, \\
    %  & = \int_{0}^{t}{(v_{ux}(\tau)+W_i(\tau)\Lambda_i(\tau) )^2\,d\tau}. \\
 \end{split}
\end{equation}

From the above equation, it is clear that the network output is  approximating $-v_{ux}(t)$. It is assumed that $v_{up}(t)\approx -v_{ux}(t)$ after a time, $t>t_l$ using the universal approximation property of the network [44]. This implies
\begin{equation}
  v_{up}(t)= -v_{ux}(t)+\epsilon(t) 
\end{equation}
where $\epsilon(t)$ is the approximation error and is bounded by a positive constant $\alpha<\infty$.
This implies that $v_{us}= \epsilon(t)$ for $t>t_l$.
So from (\ref{ss1}), we obtain
\begin{equation}
    k_3 \dot{x}_2=-k_1 x_2 - k_2 x_1 +\epsilon(t)
    \label{ss2}
\end{equation}
Using the above equation and (\ref{p1a}), we obtain the following result.
\begin{equation}
\frac{1}{1+k_3} (\dot{x}_r(t)-v_{xd}+v_{up})= \frac{-k_2}{1+k_3}x_1-\frac{-k_1}{1+k_3}x_2 +\epsilon(t)
\label{ss3}
\end{equation}
By substituting (\ref{ss3}) in (\ref{p1a}), we obtain
\begin{equation}
\left (\begin {array}{c} \dot{x}_1\\  \dot{x}_2 \end {array} \right)=\left (\begin {array}{cc} 0&1\\  \frac{-2k_2}{1+k_3}&\frac{-2k_1}{1+k_3} \end {array} \right)\left (\begin {array}{c} x_{1}\\  x_{2} \end {array} \right)+\left (\begin {array}{c} 0\\  1 \end {array} \right)\epsilon(t)
    \label{ss4}
\end{equation}
The above equation implies that the error $e_x(t)=x_2$ remains bounded as $\epsilon(t)$ is a bounded input and the matrix $\left (\begin {array}{cc} 0&1\\  \frac{-2k_2}{1+k_3}&\frac{-2k_1}{1+k_3} \end {array} \right)$  is Hurwitz for $k_1>0$, $k_2>0$ and $k_3\geq0$.\\

\textit{Remark I}: If there is a perfect learning then $\epsilon(t)=0$ and from equation (\ref{ss2}) we obtain

\begin{equation}
    s_s(t)=k_1 x_2 +k_2 x_1 +k_3 \dot{x}_2=\epsilon(t)=0
    \label{ss2ba}
\end{equation}
The above equation indicates the existence of sliding motion.

\section{Numerical Simulation Investigations}\label{sec_simulations}
Before our proposed approach is deployed into the real hardware platform, it is evaluated under a realistic simulation environment using Dronekit API in Ardupilot Software in the Loop (SITL) simulator\footnote{Simulation codes are available in this link: https://github.com/ContinualAL/PAC-with-Dronekit-SITL}. The UAV mission is to track a circular trajectory under windy conditions. The equation for the reference trajectory is given below.
\begin{equation}
    x_r(t)=x_c+A_r \cos(\frac{2\pi t}{T_{cx}})
    \label{circle_ref_x}
\end{equation}
\begin{equation}
    y_r(t)=y_c+A_r \sin(\frac{2\pi t}{T_{cy}})
    \label{circle_ref_y}
\end{equation}
where $(x_c,\,y_c)$ is the centre of the circle, $A_r$ is the radius in m and $T_c$ is the time period in $\SI{}{\s}$. Here $(x_c,\,y_c)=(0,\,0)$, $A_r = \SI{6}{\m}$ and $T_{cx} = T_{cy}= \SI{120}{\s}$. A sinusoidal signal is used to represent  the time-varying atmospheric wind gust.   The wind disturbance component is treated as an additional velocity input to the system $v_{xd}$ as in (\ref{teq11}). The expression for wind disturbance is given by
\begin{equation}
    v_{xd}=A_d+B_d \sin(2 \pi \omega_d t)
\end{equation}
The same wind disturbance is applied along inertial Y-axis. To evaluate the performance of the controller,  three different disturbance levels are considered, namely: low, medium and high. The magnitude of the constant part of the wind  $A_d \in \{-0.1,\,-0.2,\,-0.5\}$ $\SI{}{\m\per\s}$ to denote low, medium and high respectively. The magnitude of the sinusoidal component $B_d$ = $\SI{0.05}{\m\per\s}$ and the frequency $\omega_d$ = $\SI{1}{\radian\per\s}$ are the same for all disturbance levels. The PID parameters used in the simulation are given in the Table \ref{tab:smc_sim_parameters}. 

\begin{table}[b]
	\rowcolors{2}{}{def_blue}
	\centering
	\small
	\tabcolsep=0.1cm
	%	\begin{tabular}{|c|c|c|c|}
	\caption{\label{tab:smc_sim_parameters} PID parameters used for simulation.}
	\footnotesize
	\bgroup
    \def\arraystretch{1.3}
	\begin{tabular}{ll}
		\toprule
		\textbf{Control} & \pbox{20cm}{\textbf{PID parameters}} \\ 
		\textbf{$x$-position} & $k_1$=0.4, $k_2$=0.001, $k_3$=0.001, $a_m$=1.0   \\
		\textbf{$y$-position} & $k_1$=0.4, $k_2$=0.001, $k_3$=0.001, $a_m$=1.0   \\
% 		\textbf{Altitude $z$} &  $k_1$=0.35, $k_2$=0.01, $k_3$=0.65, $a_m$=0.8 \\
		\bottomrule	
	\end{tabular}
	\egroup
\end{table}

\begin{figure}[t]
\centerline{\includegraphics[width=0.53\textwidth]{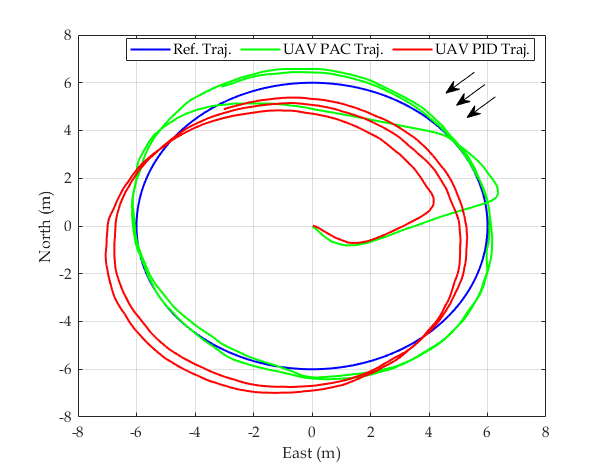}}
\caption{Trajectory tracking using PAC and PID under high level wind disturbance - the wind direction is shown by the arrow.}
\label{simulation_circle}
\end{figure}

\begin{figure}[!t]
\centerline{\includegraphics[width=0.53\textwidth]{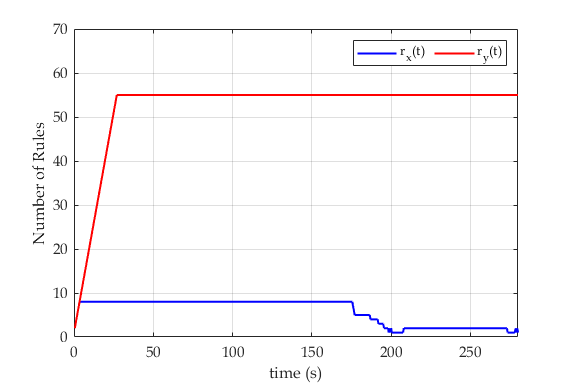}}
\caption{Number of rules generated during simulation.}
\label{simulation_rules}
\end{figure}

\begin{figure}[t]
\centerline{\includegraphics[width=0.53\textwidth]{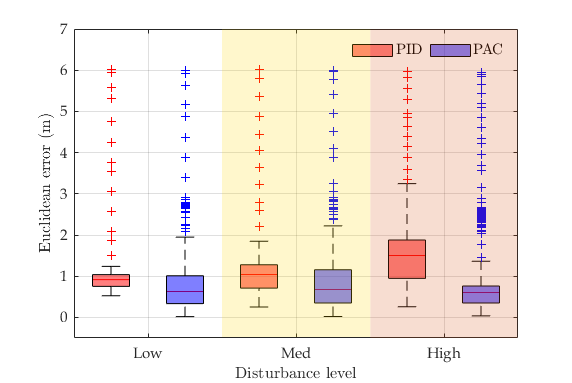}}
\caption{Simulation performance comparison.}
\label{simulation_boxplot}
\end{figure}

The  circular reference trajectory and the trajectory of the UAV for the case of PAC and PID  is shown under high level wind disturbance  in Fig. \ref{simulation_circle}. Note that the velocity input from PID is saturated at $\SI{0.8}{\m\per\s}$ and the wind velocity of $\SI{-0.5}{\m\per\s}$ is more than $50 \%$ of the control input. The blue curve denotes the UAV trajectory using PAC and is closer to the reference trajectory when compared to the PID controller denoted by the green curve. The PAC  has a self organizing property to autonomously adjust the number of fuzzy rules denoted by $\big(r_x(t),\,r_y(t)\big)$.  Initially, the output of the PALM network is generated from  a single rule and then it grows depending on the expected bias and variance as given in (\ref{eq:add_R}) and (\ref{eq:prune_R}). The rule evolution for the case of  circular trajectory tracking with high wind disturbance is  shown in Fig. \ref{simulation_rules}. To compare the trajectory tracking performance, Root Mean Square Error (RMSE)  is used and calculated as:
\begin{equation}
    \label{eq:rmse}
    \text{RMSE}=\sqrt{\frac{1}{N}\sum_{k=1}^{N} \bar{e}(k)^{\rm T}\bar{e}(k)}
\end{equation}
where $\bar{e}(k)=[e_x(k),\,e_y(k),\,e_z(k)]^{\rm T}$ and $k$ is the sampling instant. The total number of samples is denoted by $N=\frac{T_f}{t_s}$, with $T_f$ is the total time and $\SI{0.01}{\s}$ is the sampling interval. 
The comparison of RMSE values $(\SI{}{\m})$ is presented in Table \ref{tab:RMSE_sim}. For the case of low disturbance, the performance of the PID controller and PAC is almost the same. When the disturbance input is increased from low to high, the RMSE of the PID controller increases by $\SI{0.55}{\m}$. Meanwhile, the RMSE of PAC increases only by $\SI{0.15}{\m}$. The distribution of Euclidean error as given in the below equation, is further analyzed.
\begin{equation}
    \varepsilon_{xy}(k)=\sqrt{e_x(k)^2+e_y(k)^2}
\end{equation}
The statistical boxplot of Euclidean error is shown in Fig. \ref{simulation_boxplot}.

\begin{table}[b]
	\rowcolors{2}{}{def_blue}
	\centering
	\small
	\tabcolsep=0.1cm
	%	\begin{tabular}{|c|c|c|c|}
	\caption{\label{tab:RMSE_sim} RMSE values of  circle trajectory simulation.}
	\footnotesize
	\bgroup
    \def\arraystretch{1.3}
	\begin{tabular}{lll}
		\toprule
		\textbf{Disturbance level} & \pbox{20cm}{\textbf{ PID}} &
		\pbox{20cm}{\textbf{ PAC}}\\ 
		\hline
		\textbf{Low} & $1.1033$ & $1.0610$   \\
		\textbf{Medium} &  $1.2116$ & $1.1629$  \\
		\textbf{High} & $1.6516$ & $1.2093$ \\
		\bottomrule	
	\end{tabular}
	\egroup
\end{table}

When the disturbance level is low, the trajectory tracking error of PAC and PID controller is quite similar, although PAC shows a slightly lower mean value of $\varepsilon_{xy}$ but higher variance. Meanwhile, if the disturbance is higher, the mean and variance of  $\epsilon_{xy}$  for the case of PAC are significantly lower than that of the PID controller.  It is evident that by having the appropriate equivalent input learned by the neuro-fuzzy part of the PAC  is  able to reduce the effect of wind disturbances.  The numerical simulation results indicate that the PAC attenuates the wind disturbances higher than the conventional PID controller.

\begin{figure*}[!t]
\centerline{\includegraphics[scale=0.35]{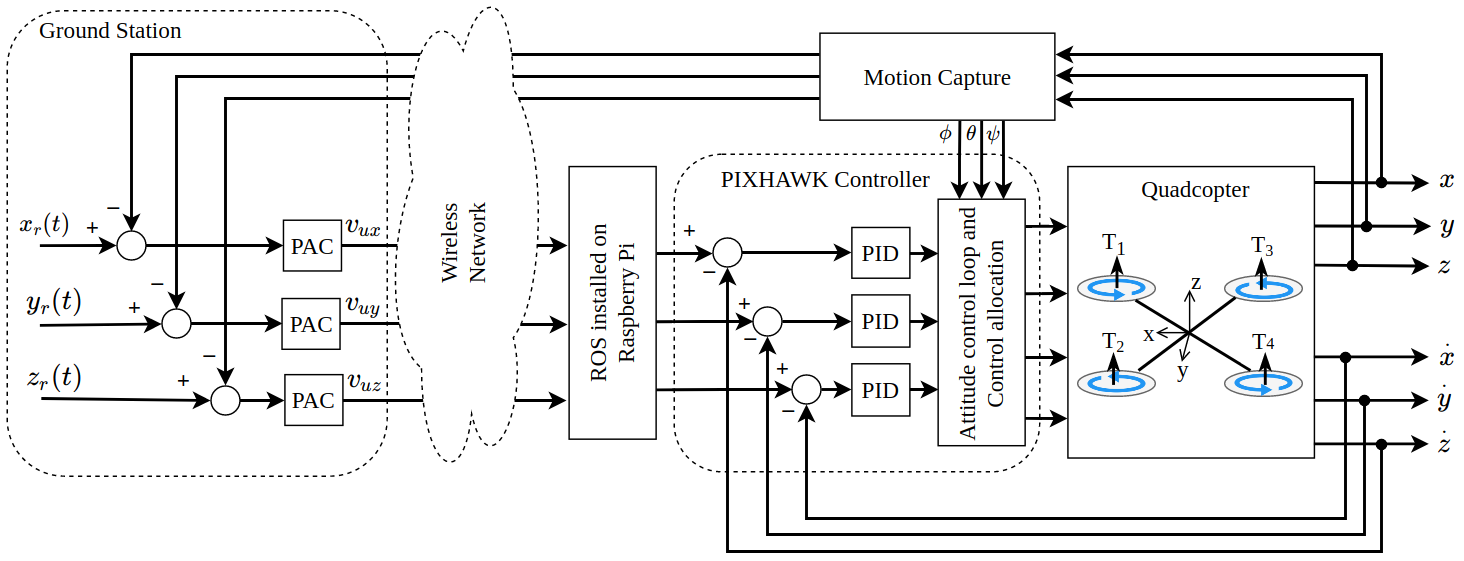}}
\caption{Architecture of the controller implementation.}
\label{sysblock}
\end{figure*}

\color{black}
\section{Experimental Results}\label{sec_experiments}

To evaluate the tracking and disturbance rejection properties of the PAC, several experiments are conducted using a quadrotor UAV\footnote[2]{Experimental video is available in: \url{https://youtu.be/-nGhMzvB7oE}}. The details of the experimental hardware used and the results of trajectory tracking, disturbance rejection are explained in the following two subsections. 

\subsection{Experimental Setup and Methodology}

\begin{figure}[h]
\centerline{\includegraphics[width=0.475\textwidth]{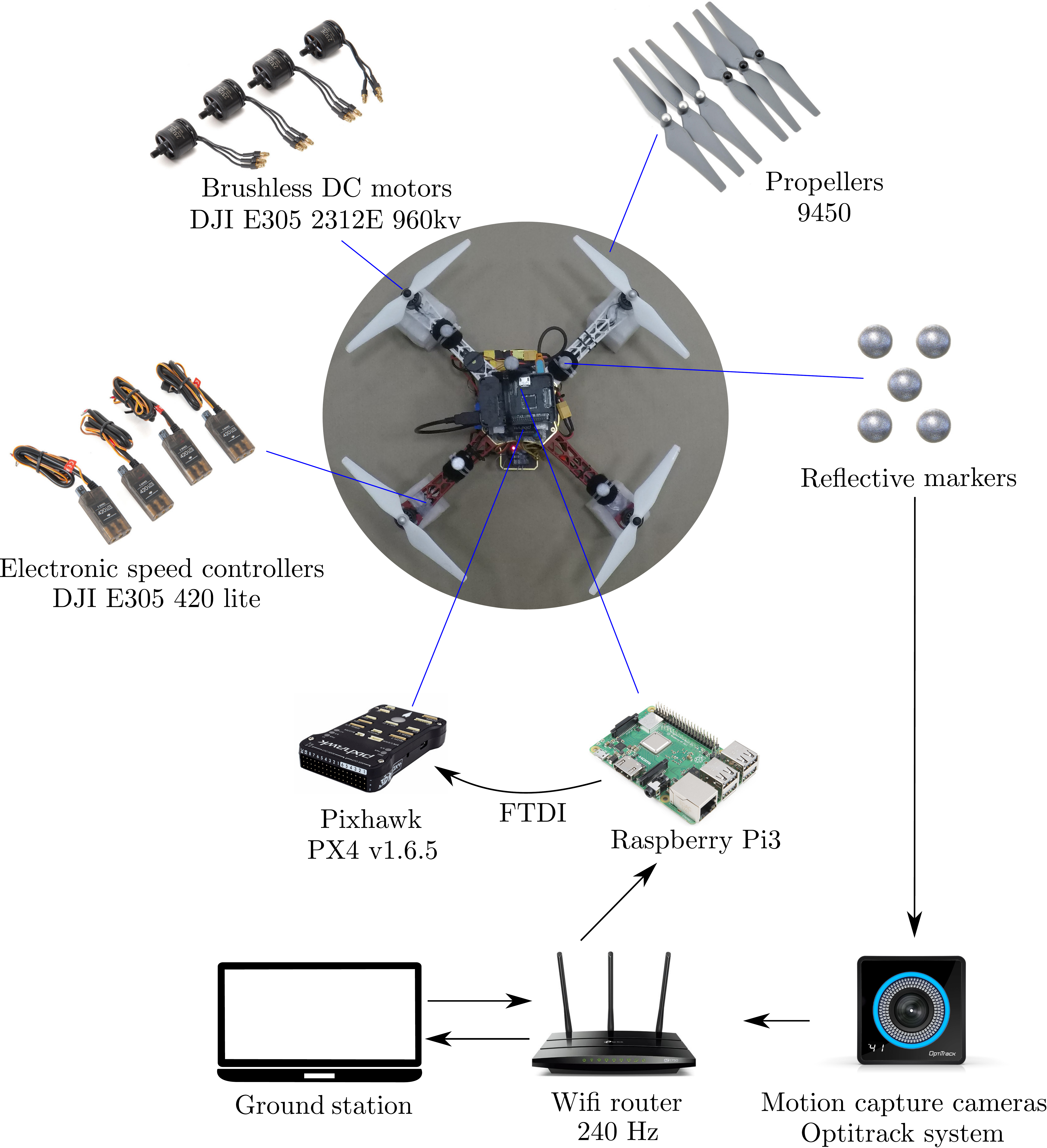}}
\caption{Experimental setup: Quadrotor and its components.}
\label{exp_setup1}
\end{figure}

The experiments are conducted in an indoor environment using a quadrotor UAV of dimension $\SI{25}{\cm}$ between the rotors with a propeller of diameter 9 inches. The UAV has onboard Pixhawk autopilot hardware interfaced to a Raspberry-Pi3 computer. The position and orientation of the UAV are precisely measured using the OptiTrack vision system having eight infrared cameras. The position and orientation data are transmitted over a wifi network to a Ground Control Station (GCS) computer at a rate of $\SI{240}{\Hz}$. The control algorithm is executed in MATLAB environment inside the GCS and the control input is transmitted to the Raspberry-Pi3 computer via wifi network. The Raspberry-Pi3 computer publishes the reference velocity inputs to the Pixhawk via Robot Operating System (ROS) platform. For the inner loop velocity tracking control, a PID controller is used. The architecture of the controller implementation is shown in Fig. \ref{sysblock} and our experimental setup is depicted in Fig. \ref{exp_setup1}.

\subsection{Results and Analysis}

\begin{figure*}[!t]
\centerline{\includegraphics[scale=0.34]{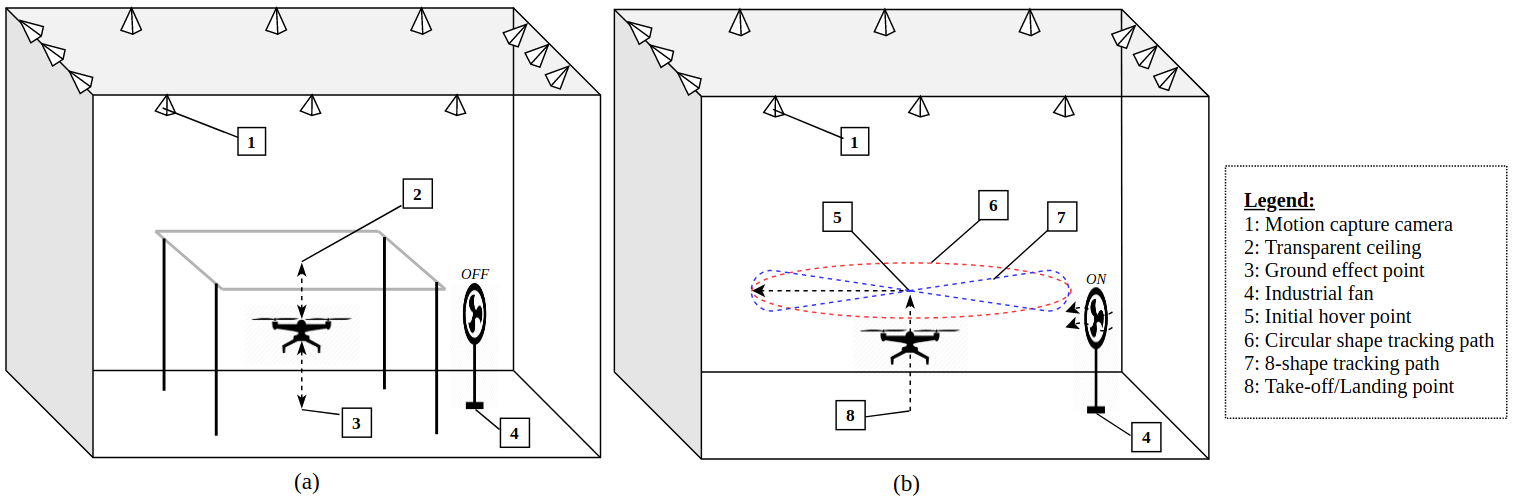}}
\caption{Experimental setup in motion capture laboratory: (a) Setup to be used for the experiment in CASE C including the transparent ceiling on top of the UAV and ground-effect; (b) Setup for circular, 8-shape tracking  CASE A (without disturbance) and B to test the UAV with wind gust disturbance blown by an industrial fan.}
\label{sysblock_exp}
\end{figure*}

The performance of PAC is experimentally evaluated for tracking various challenging trajectories as well as interaction scenarios. The typical trajectories considered here are circular and eight shaped as they actuate the baseline system dynamics more when compared to a straight line or rectangular trajectories. Moreover, when UAV is navigating in a restricted environment having obstacles, the admissible trajectories will be of circular or eight shaped in nature. For the interaction case, close proximity effects are tested, including ground and ceiling effects. The experiments are conducted without any disturbance input and also for the case of wind disturbances generated using a fan. All these cases are illustrated in Fig. \ref{sysblock_exp}. The proposed controller PAC  is compared against the baseline PID to demonstrate the advantage of online adaptation law as well as an elastic network structure. In the presence of wind disturbances, ground and ceiling effects, the evaluations are quantified using RMSE values, which is defined in (\ref{eq:rmse})
%
%
%
% \begin{equation}
%     RMSE=\sqrt{\frac{1}{N}\sum_{k=1}^{N} \bar{e}(k)^{\rm T}\bar{e}(k)},
% \end{equation}
%
%
 The PID parameters for comparison are given in Table \ref{tab:smcspecs}.
\begin{table}[b!]
	\rowcolors{2}{}{def_blue}
	\centering
	\small
	\tabcolsep=0.1cm
	%	\begin{tabular}{|c|c|c|c|}
	\caption{\label{tab:smcspecs} PID parameters used for experiments.}
	\footnotesize
	\bgroup
    \def\arraystretch{1.3}
	\begin{tabular}{ll}
		\toprule
		\textbf{Control} & \pbox{20cm}{\textbf{PID parameters}} \\ 
		\textbf{$x$-position} & $k_1$=1.0, $k_2$=0.001, $k_3$=0.0, $a_m$=0.8   \\
		\textbf{$y$-position} & $k_1$=1.0, $k_2$=0.001, $k_3$=0.0, $a_m$=0.8   \\
		\textbf{Altitude $z$} &  $k_1$=0.35, $k_2$=0.01, $k_3$=0.65, $a_m$=0.8 \\
		\bottomrule	
	\end{tabular}
	\egroup
\end{table}

Another critical situation arises when the UAV operates in close proximity to the ceiling or the ground. The interactions between the propeller wake and the ceiling or ground act as input disturbance to the system. Experiments are conducted for the case of tracking a square wave with upper and lower limits set close to ceiling and ground, respectively. Each of the above-mentioned experimental cases is explained below in detail.

\begin{figure}[h]
\centerline{\includegraphics[height=6.5cm,width=9.5cm]{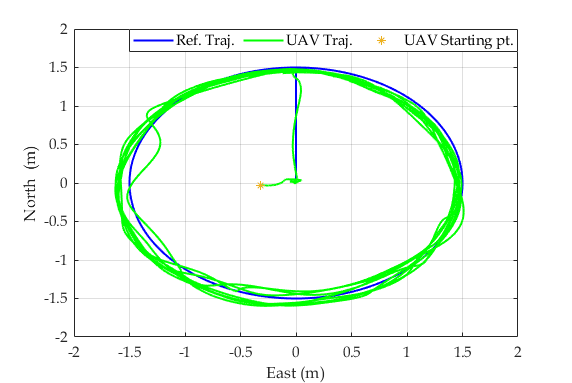}}
\caption{Circle shaped trajectory tracking using PAC.}
\label{figs11}
\end{figure}

\begin{figure}[h]
\centerline{\includegraphics[height=6.7cm,width=9.7cm]{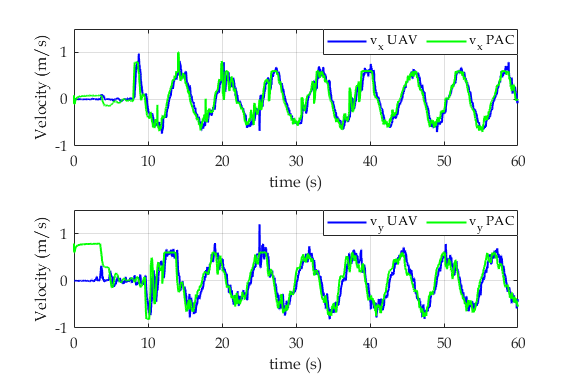}}
\caption{Velocity command generated by PAC and UAV velocity for circle shaped trajectory tracking.}
\label{figs12}
\end{figure}

\begin{figure}[h]
\centerline{\includegraphics[height=7cm,width=10cm]{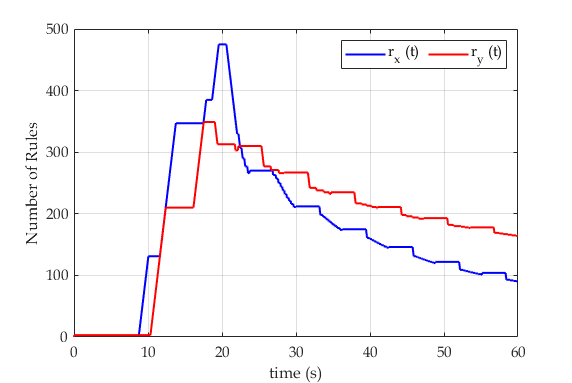}}
\caption{Rule evolution for circle shaped trajectory tracking.}
\label{figs13}
\end{figure}

\begin{figure}[h]
\centerline{\includegraphics[height=7cm,width=10cm]{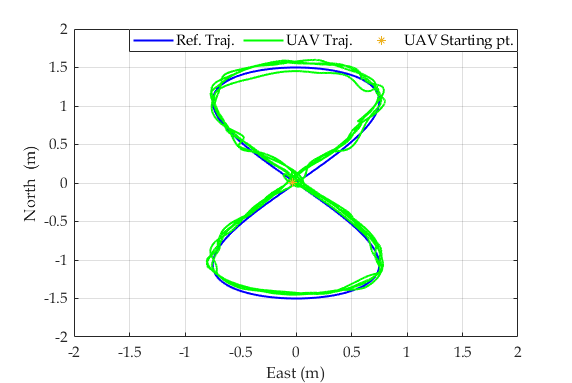}}
\caption{Eight shaped trajectory tracking using PAC }
\label{figs21}
\end{figure}

\begin{figure}[h]
\centerline{\includegraphics[height=7cm,width=10cm]{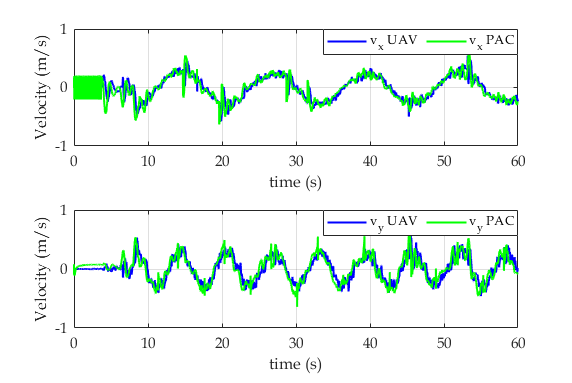}}
\caption{Velocity command generated by PAC and UAV velocity for eight shaped trajectory tracking }
\label{figs22}
\end{figure}

\begin{figure}[h]
\centerline{\includegraphics[height=6.6cm,width=9.6cm]{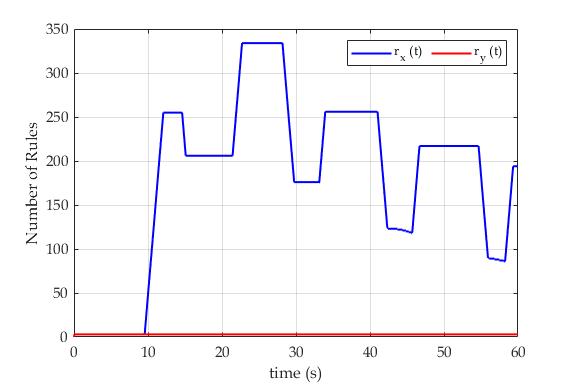}}
\caption{Rule evolution for eight shaped trajectory tracking.}
\label{figs23}
\end{figure}

\textit{CASE A: Circular and eight shaped trajectory tracking.}\\
A circle of diameter $\SI{3}{\m}$ is taken as the reference trajectory as given in (\ref{circle_ref_x}) and (\ref{circle_ref_y}). Here $(x_c,\,y_c)=(0,\,0)$ $\SI{}{\m}$, $A_r=1.5$ $\SI{}{\m}$ and $T_c=\SI{10}{\s}$. The trajectory followed by the UAV and the reference circular trajectory is shown in Fig. \ref{figs11}. The UAV follows a straight line initially to intercept the circle at $(1.5, 0)$ $\SI{}{\m}$. The velocity command generated by PAC and the velocity of UAV is shown in Fig. \ref{figs12}. A constant non-zero velocity reference is generated by PAC initially to follow a straight line to intercept the circle. A sinusoidal velocity command is generated by PAC within the saturation velocity of $\SI{1}{\m\per\s}$ along any axis. The evolution of rules with respect to time for the generation of reference $v_x$ and $v_y$ using PAC is shown in Fig. \ref{figs13}. The  number of  rules  $\big(r_x(t),\,r_y(t)\big)$ is very small for time $t \leq \SI{10}{\s}$ as a straight line is followed. The number of rules increases during the initial transition from the straight line to the circular trajectory. As the UAV continues to track the circular trajectory, the number of rules decreases for $t \geq \SI{20}{\s}$. The neuro-fuzzy network adapts to the circular trajectory and as a result of that, the excess number of rules are pruned for $t \geq  \SI{20}{\s}$. The number of rules generated is maximum during the transition from the straight line to the circular trajectory, i.e. for $10<t< \SI{20}{\s}$.
The eight shaped reference trajectory is generated by \eqref{circle_ref_x} and \eqref{circle_ref_y}
% %
% \begin{equation}
%     x_r(t)=x_c+A_x sin(\frac{2\pi t}{T_x}),
%     \label{81}
% \end{equation}
% %
% \begin{equation}
%     y_r(t)=y_c+A_y sin(\frac{2\pi t}{T_y}),
%     \label{82}
% \end{equation}
% %
where the time periods $(T_{cx})$ and $(T_{cy})$are  $\SI{20}{\s}$ and  $\SI{10}{\s}$ respectively. The amplitudes of the sinusoidal signals $A_r$ are  $\SI{1.5}{\m}$ and $\SI{0.75}{\m}$ respectively. The centre coordinates are $(x_c,\,y_c)=(0,\,0)$ $\SI{}{\m}$. The reference trajectory and the response of the UAV attained with PAC are shown in Fig. \ref{figs21}. The corresponding velocity command and UAV velocity profiles are shown in Fig. \ref{figs22}. The PAC generates a higher frequency velocity command along Y-axis when compared to X-axis in concurrence with \eqref{circle_ref_x} and \eqref{circle_ref_y}. The rule evolution for eight shaped trajectory tracking is shown in Fig. \ref{figs23}. The number of rules $r_y(t)$ is smaller when compared to $r_x(t)$ as the amplitude and the time period of the sinusoidal velocity along Y-axis is lower when compared to that of X-axis. The difference in amplitude and time period makes the UAV to travel higher distance along X-axis when compared to that along Y-axis as seen from Fig. \ref{figs21}.

\begin{figure}[t]
\centerline{\includegraphics[height=7cm,width=10cm]{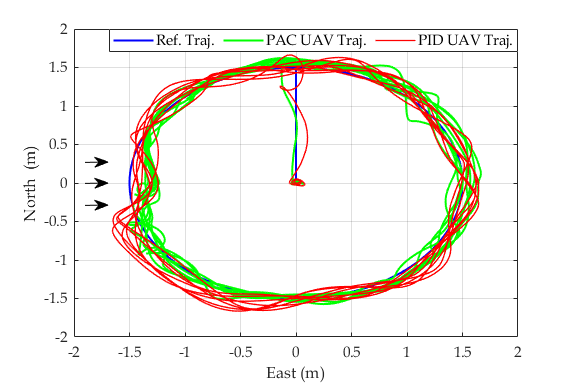}}
\caption{Comparison between PAC and PID for circle shaped trajectory tracking with wind disturbance - the wind direction is shown by the arrow.}
\label{figs31}
\end{figure}

\begin{figure}[t]
\centerline{\includegraphics[height=7cm,width=10cm]{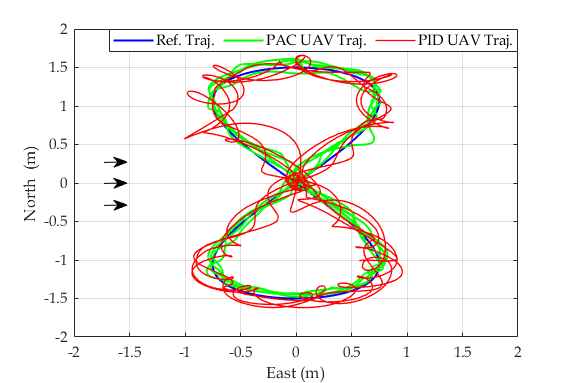}}
\caption{Comparison between PAC and PID for eight shaped trajectory tracking with wind disturbance - the wind direction is shown by the arrow.}
\label{figs32}
\end{figure}

\textit{CASE B:  Trajectory tracking under wind disturbances.}\\
The effect of output disturbance on the performance of PAC is verified by generating a wind disturbance using a fan. A comparison is provided with PID for tracking circular and eight shaped trajectories mentioned in CASE A. The trajectories obtained using PAC and PID are shown in Fig. \ref{figs31} for the case of tracking the circular trajectory. The direction of the wind is shown by the arrow. It can be observed that with PAC, the deviation from the trajectory is less when compared to that of using PID. For the case of tracking eight shaped trajectories with wind disturbance, the responses of PAC and PID are shown in Fig. \ref{figs32}. The RMSE values shown in Table \ref{table2} indicate that PAC excels PID in rejecting wind disturbances.

\begin{table}
\rowcolors{2}{}{def_blue}
\centering
\tabcolsep=0.1cm
\caption{\label{table2} Summary of RMSE values of the experiments.}
\bgroup
\def\arraystretch{1.3}
\begin{tabular}{c|cc|cc}
\toprule
\hline 
\multirow{2}{*}{\textbf{UAV Mission}} & \multicolumn{2}{c|}{\textbf{RMSE no wind $(\SI{}{\m})$ }} & \multicolumn{2}{c}{\textbf{RMSE with wind $(\SI{}{\m})$}}\tabularnewline
\cline{2-5} 
 & \textbf{PID} & \textbf{PAC} & \textbf{PID} & \textbf{PAC}\tabularnewline
\hline
\hline
Circular-shape & $0.2349 $ & $0.1515 $ & $0.2602$ & ${0.1768}$\tabularnewline
\hline 
8-shape & $0.1255 $ & $0.1216 $ & $0.1945$ & ${0.1254}$\tabularnewline
\hline 
Altitude with interaction & $0.0987$ & ${0.0859}$ & N/A & N/A\tabularnewline
\hline 
\bottomrule	 
\end{tabular}
\egroup
\end{table}

\begin{figure}[b]
\centerline{\includegraphics[height=6.5cm,width=9.5cm]{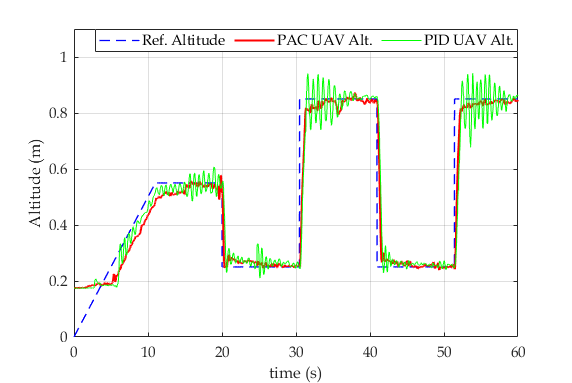}}
\caption{Comparison between PAC and PID for altitude tracking  in the presence of ceiling and ground effects.}
\label{figsc1}
\end{figure}

\textit{CASE C: Altitude tracking in the presence of ceiling and ground interaction effects.}\\
Considering an inspection of  surrounding with UAVs [45], the interaction between the propeller wake and ceiling/ground generates additional forces that can be treated as input disturbance acting on the system. A square wave of amplitude $\SI{0.6}{\m}$ and time period $\SI{21}{\s}$ is given as a reference trajectory for altitude tracking. The responses of PAC and PID are given in Fig. \ref{figsc1}. The reference altitude varies from $0.25$ - $0.85$ $\SI{}{\m}$. As observed in Fig. \ref{figsc1}, the input disturbance due to the ceiling effect is more prominent when compared to the disturbance created by the ground effect. The PID  control has a peak overshoot of $10.5$ $\%$ while tracking  $\SI{0.85}{\m}$ altitude reference, whereas PAC has only $1.8$ $\%$ peak overshoot. In the presence ground effect, it is hard to remain closer to the reference altitude due to the additional upward force generated by the rotors. For a reference altitude of $\SI{0.25}{\m}$, the altitude envelope of PAC is in between $0.241$ - $0.301$ $\SI{}{\m}$ when compared to $0.213$ - $0.332$ $\SI{}{\m}$ achieved by PID. The RMSE values are given in Table \ref{table2}, with PAC having the lowest value. Thus PAC is more effective than PID in attenuating input disturbances.

\section{Discussion}\label{sec_discussion}
The combined sliding surface and the PALM network act as an SMC. From the performance evaluations in both numerical simulations and experiments, it is observed that the PAC  outperforms PID in the presence of disturbances. From the the simulation boxplot (Fig. \ref{simulation_boxplot}) and the RMSE values shown in Table \ref{tab:RMSE_sim} and Table \ref{table2}, the PAC performs consistently better than the baseline PID controller. These results validate the importance of the proposed concept of combining the autonomous learning machine as the equivalent input of the SMC. Unlike the conventional SMC, where the knowledge on the upper bound on the disturbance input is required, here, the structure of the PALM network is adapted autonomously to compensate for the disturbances. The plot of wind disturbance and the control input generated by the PALM network of PAC provided in the supplementary material shows that the control action provided by the PALM network opposes the wind disturbance. In effect, the error dynamics is being pushed towards the sliding surface. The  elastic structure and online parameter adaptation of PAC enable the controller to adapt under different external disturbances. The structure and parameters of the PID controller are static, restricting its attenuation capabilities for disturbances with different  amplitude and bandwidth.

The autonomous learning machine used in this work has an elastic structure to tackle the issue of selecting an appropriate  network structure for the problem. The network structure is self-constructed online, depending on the mission complexity. The number of rules of the PALM network  depends on the rule growing and pruning criteria given in equation (\ref{eq:add_R}) and (\ref{eq:prune_R}) \footnote{Detailed analysis of the number of rules generated for a different setting of the rule pruning parameters is provided in the supplementary material.}. The membership functions of the PAC is based on the hyper-plane distance function. It evolves to the desired structure following the bias and variance trade-off, starting from a simple structure. 

The experiments performed for tracking circular and eight shaped trajectories and altitude hold under ceiling and ground effects are performed with  fixed parameters for the sliding surface as given in Table \ref{tab:smcspecs}. As a comparison for experimental conditions, the statistical error analysis is provided in the supplementary material for the case of ceiling effect. The proposed PAC is compared with the standard PID controller and also an optimization based Nonlinear Model Predictive Controller (NMPC). In close proximity, particularly for the rotor radius's range, both of the controllers have large tracking  errors when compared to PAC.
The PALM network part of the PAC generates a different number of rules for each of these experimental tasks and compensate for disturbances without any further parameter tuning. The autonomous evolving feature of the PAC alleviates the designer's effort to tune the network structure or the parameters of the sliding surface for different experimental scenarios. 

% - simulation rmse and boxplot
% - experiment rmse
% - why pac can perform better?
% 1) has ueq learned online
% 2) has elastic structure
% - The benefit of PAC is able to work from only one rule, does not need any pre-defined structure.

% \textcolor{blue}{The presented experiments are based on the offboard localization case where the accuracy of the body pose is precise. However, as compared to our former result, which uses baseline MPC \cite{kocer2019aerial}, the performance of the presented approach is quite effective to compensate for the aerodynamic interactions and disturbances. On the other hand, in an onboard localization case, the system needs a particular deployment of sensors. For an outdoor implementation, the system can be endowed by a real-time kinematic GPS which can provide centimeter-level accuracy. Since GPS is not a viable option for an indoor application, visual-inertial odometry could be leveraged for the UAV. If the application area is lack of features and proper illumination, laser-based sensors (e.g., Lidar) can be added to the system. In each case, the effect on the system could be different. The major effects can be: (i) noise; (ii) drift; (iii) loop frequency. In these cases, the controller might need practical adjustments but analysis of them is beyond the presented scope.}

\color{black}
\section{Conclusion}\label{sec_conclusion}

This paper presents an autonomous controller consisting of a self organized neuro-fuzzy structure and a sliding surface with a structure of a PID controller. After defining the mathematical model, the challenges are identified for the problem that arises during the interaction phase, as well as tracking while flying through the wind gust disturbance case. Afterward, the proposed evolving neuro-fuzzy structure is introduced. Stability analysis is provided to show that the tracking error remains bounded in the presence of disturbance input. The experiments showed that the system performance under several different missions is recovered when the autonomous controller is applied. The proposed controller eases the design and implementation process as it does not require any further tuning for a different task, a precise system model, and exact information about the environment. For the sake of reproducible research, the link to our simulation code and experimental video is also provided.

An interesting extension of this work would be the application to more challenging scenarios such as trajectory and target tracking in an outdoor environment [46]. In this context, reinforcement learning-based approaches (e.g., [47]-[48]) can be explored within interaction scenarios for the field tests.

\section*{Acknowledgment}
The research work was supported by the Nanyang Technological University internal grant for the development of a large VTOL research platform.

%\bibliography{References/bib,References/IEEEabrv}
%\bibliographystyle{References/IEEEtran}

 \section*{References}

[1] A. C. Woods and H. M. La, "A novel potential field controller for use
on aerial robots," IEEE Transactions on Systems, Man, and Cybernetics:
Systems, vol. 49, no. 4, pp. 665-676, 2017.

[2] S. Islam, P. X. Liu, A. El Saddik, R. Ashour, J. Dias, and L. D. Seneviratne,
"Artificial and virtual impedance interaction force reflection-based
bilateral shared control for miniature unmanned aerial vehicle," IEEE
Transactions on Industrial Electronics, vol. 66, no. 1, pp. 329-337, 2019.

[3] K. Dorling, J. Heinrichs, G. G. Messier, and S. Magierowski, "Vehicle
routing problems for drone delivery,"IEEE Transactions on Systems,
Man, and Cybernetics: Systems, vol. 47, no. 1, pp. 70-85, 2016.

[4] X. He, G. Kou, M. Calaf, and K. K. Leang, "In-ground-effect modeling
and nonlinear-disturbance observer for multirotor unmanned aerial vehicle
control," Journal of Dynamic Systems, Measurement, and Control,
vol. 141, no. 7, 2019

[5] S. J. Portugal, T. Y. Hubel, J. Fritz, S. Heese, D. Trobe, B. Voelkl,
S. Hailes, A. M. Wilson, and J. R. Usherwood, "Upwash exploitation and
downwash avoidance by flap phasing in ibis formation flight," Nature,
vol. 505, no. 7483, pp. 399-402, 2014.

[6] A. E. Jimenez-Cano, P. J. Sanchez-Cuevas, P. Grau, A. Ollero, and
G. Heredia, "Contact-based bridge inspection multirotors: Design, modeling,
and control considering the ceiling effect," IEEE Robotics and
Automation Letters, vol. 4, no. 4, pp. 3561-3568, 2019.

[7] B. B. Kocer, V. Kumtepeli, T. Tjahjowidodo, M. Pratama, A. Tripathi,
G. S. G. Lee, and Y. Wang, "Uav control in close proximities-ceiling
effect on battery lifetime," in 2019 2nd International Conference on
Intelligent Autonomous Systems (ICoIAS). IEEE, 2019, pp. 193-197.

[8] S. Gao, C. Di Franco, D. Carter, D. Quinn, and N. Bezzo, "Exploiting
ground and ceiling effects on autonomous uav motion planning," in
2019 International Conference on Unmanned Aircraft Systems (ICUAS).
IEEE, 2019, pp. 768-777.

[9] C. Powers, D. Mellinger, A. Kushleyev, B. Kothmann, and V. Kumar,
"Influence of aerodynamics and proximity effects in quadrotor flight,"
in Experimental robotics. Springer, 2013, pp. 289-302.

[10] B. B. Kocer, T. Tjahjowidodo, and G. G. L. Seet, "Centralized predictive
ceiling interaction control of quadrotor VTOL UAV," Aerospace Science
and Technology, vol. 76, pp. 455-465, 2018.

[11] T. Gao, S. Yin, J. Qiu, H. Gao, and O. Kaynak, "A partial least squares
aided intelligent model predictive control approach," IEEE Transactions
on Systems, Man, and Cybernetics: Systems, vol. 48, no. 11, pp. 2013-
2021, 2017.

[12] L. Besnard, Y. B. Shtessel, and B. Landrum, "Control of a quadrotor
vehicle using sliding mode disturbance observer," in 2007 American
Control Conference. IEEE, 2007, pp. 5230-5235.

[13] K. Alexis, G. Nikolakopoulos, and A. Tzes, "Switching model predictive
attitude control for a quadrotor helicopter subject to atmospheric disturbances,"
Control Engineering Practice, vol. 19, no. 10, pp. 1195-1207,
2011.

[14] B. B. Kocer, T. Tjahjowidodo, and G. G. L. Seet, "Model predictive
UAV-tool interaction control enhanced by external forces," Mechatronics,
vol. 58, pp. 47-57, 2019.

[15] A. Das, K. Subbarao, and F. Lewis, "Dynamic inversion with zerodynamics
stabilisation for quadrotor control," IET control theory $\&$
applications, vol. 3, no. 3, pp. 303-314, 2009.

[16] E. J. Smeur, G. C. de Croon, and Q. Chu, "Cascaded incremental
nonlinear dynamic inversion for MAV disturbance rejection," Control
Engineering Practice, vol. 73, pp. 79-90, 2018.

[17] Z. T. Dydek, A. M. Annaswamy, and E. Lavretsky, "Adaptive control
of quadrotor UAVs: A design trade study with flight evaluations," IEEE
Transactions on control systems technology, vol. 21, no. 4, pp. 1400-
1406, 2013.

[18] Y. Zou and Z. Meng, "Immersion and invariance-based adaptive controller
for quadrotor systems," IEEE Transactions on Systems, Man, and
Cybernetics: Systems, 2018.

[19] S. Islam and X. P. Liu, "Robust sliding mode control for robot manipulators,"
IEEE Transactions on Industrial Electronics, vol. 58, no. 6, pp.
2444-2453, 2011.

[20] N. Wang, S. Su, M. Han, and W. Chen, "Backpropagating constraintsbased
trajectory tracking control of a quadrotor with constrained actuator
dynamics and complex unknowns," IEEE Transactions on Systems, Man,
and Cybernetics: Systems, vol. 49, no. 7, pp. 1322-1337, July 2019.

[21] L. Besnard, Y. B. Shtessel, and B. Landrum, "Quadrotor vehicle control
via sliding mode controller driven by sliding mode disturbance observer,"
Journal of the Franklin Institute, vol. 349, no. 2, pp. 658-684, 2012.

[22] B. Tian, J. Cui, H. Lu, Z. Zuo, and Q. Zong, "Adaptive finite-time
attitude tracking of quadrotors with experiments and comparisons," IEEE
Transactions on Industrial Electronics, 2019.

[23] H. Rios, R. Falcon, O. A. Gonzalez, and A. Dzul, "Continuous slidingmode
control strategies for quadrotor robust tracking: real-time application,"
IEEE Transactions on Industrial Electronics, vol. 66, no. 2, pp.
1264-1272, 2019.

[24] D. Cabecinhas, R. Cunha, and C. Silvestre, "A globally stabilizing
path following controller for rotorcraft with wind disturbance rejection,"
IEEE Transactions on Control Systems Technology, vol. 23, no. 2, pp.
708-714, March 2015.

[25] B. Xiao and S. Yin, "A new disturbance attenuation control scheme for
quadrotor unmanned aerial vehicles," IEEE Transactions on Industrial
Informatics, vol. 13, no. 6, pp. 2922-2932, December 2017.

[26] X. Lyu, J. Zhou, H. Gu, Z. Li, S. Shen, and F. Zhang, "Disturbance
observer based hovering control of quadrotor tail-sitter VTOL uavs using
H-infinity synthesis," IEEE Robotics and Automation Letters, vol. 3, no. 4,
pp. 2910-2917, October, 2018.

[27] Z. Dydek, A. Annaswamy, and E. Lavretsky, "Combined/composite
adaptive control of a quadrotor UAV in the presence of actuator
uncertainty," in AIAA Guidance, Navigation, and Control Conference,
2010, p. 7575.

[28] L. Doitsidis, K. P. Valavanis, N. C. Tsourveloudis, and M. Kontitsis,
"A framework for fuzzy logic based UAV navigation and control,"
in IEEE International Conference on Robotics and Automation, 2004.
Proceedings. ICRA-04. 2004, vol. 4. IEEE, 2004, pp. 4041-4046.

[29] V. Artale, M. Collotta, G. Pau, and A. Ricciardello, "Hexacopter trajectory
control using a neural network," in AIP Conference Proceedings,
vol. 1558, no. 1. AIP, 2013, pp. 1216-1219.

[30] B. Xu, "Composite learning finite-time control with application to
quadrotors," IEEE Transactions on Systems, Man, and Cybernetics:
Systems, vol. 48, no. 10, pp. 1806-1815, Oct 2018.

[31] S. Kurnaz, O. Cetin, and O. Kaynak, "Adaptive neuro-fuzzy inference
system based autonomous flight control of unmanned air vehicles,"
Expert Systems with Applications, vol. 37, no. 2, pp. 1229-1234, 2010.

[32] F. Santoso, M. A. Garratt, and S. G. Anavatti, "Hybrid pd-fuzzy and
pd controllers for trajectory tracking of a quadrotor unmanned aerial
vehicle: Autopilot designs and real-time flight tests," IEEE Transactions
on Systems, Man, and Cybernetics: Systems, pp. 1-13, 2019.

[33] F. Santoso, M. A. Garratt, S. G. Anavatti, and I. Petersen, "Robust hybrid
nonlinear control systems for the dynamics of a quadcopter drone," IEEE
Transactions on Systems, Man, and Cybernetics: Systems, pp. 1-13,
2018.

[34] E. Kayacan and R. Maslim, "Type-2 fuzzy logic trajectory tracking
control of quadrotor VTOL aircraft with elliptic membership functions,"
IEEE/ASME Transactions on Mechatronics, vol. 22, no. 1, pp. 339-348,
2016.

[35] M. M. Ferdaus, M. Pratama, S. G. Anavatti, M. A. Garratt, and Y. Pan,
"Generic evolving self-organizing neuro-fuzzy control of bio-inspired
unmanned aerial vehicles," IEEE Transactions on Fuzzy Systems, 2019.

[36] M. M. Ferdaus, M. Pratama, S. G. Anavatti, M. A. Garratt, and
E. Lughofer, "PAC: A novel self-adaptive neuro-fuzzy controller for
micro aerial vehicles," Information Sciences, vol. 512, pp. 481-505,
2020.

[37] K. Harikumar, J. Senthilnath, and S. Sundaram, "Mission aware motion
planning (MAP) framework with physical and geographical constraints
for a swarm of mobile stations," IEEE Transactions on Cybernetics,
vol. 50, no. 3, pp. 1209-1219, March 2020.

[38] M. Chen, S. Xiong, and Q. Wu, "Tracking flight control of quadrotor
based on disturbance observer," IEEE Transactions on Systems, Man,
and Cybernetics: Systems, pp. 1-10, 2019.

[39] N. Wang, S.-F. Su, M. Han, and W.-H. Chen, "Backpropagating
constraints-based trajectory tracking control of a quadrotor with constrained
actuator dynamics and complex unknowns," IEEE Transactions
on Systems, Man, and Cybernetics: Systems, vol. 49, no. 7, pp. 1322-
1337, 2018.

[40] J. Xu, P. Shi, C.-C. Lim, C. Cai, and Y. Zou, "Reliable tracking
control for under-actuated quadrotors with wind disturbances," IEEE
Transactions on systems, man, and cybernetics: systems, vol. 49, no. 10,
pp. 2059-2070, 2018.

[41] A. Ashfahani and M. Pratama, "Autonomous deep learning: Continual
learning approach for dynamic environments," in Proceedings of the
2019 SIAM International Conference on Data Mining. SIAM, 2019,
pp. 666-674.

[42]  M. M. Ferdaus, M. Pratama, S. G. Anavatti, and M. A. Garratt, "PALM:
An incremental construction of hyperplanes for data stream regression,"
IEEE Transactions on Fuzzy Systems, vol. 27, no. 11, pp. 2115-2129,
2019.

[43] J. A. Farrel and M. M. Polycarpou, Adaptive Approximation Based
Control. Wiley-Interscience, 2006.

[44]  K. Hornik, M. Stinchcombe, and H. White, "Universal approximation of
an unknown mapping and its derivatives using multilayer feedforward
networks," Neural networks, vol. 3, no. 5, pp. 551-560, 1990.

[45]  B. B. Kocer, T. Tjahjowidodo, M. Pratama, and G. G. L. Seet,
"Inspection-while-flying: An autonomous contact-based nondestructive
test using UAV-tools," Automation in Construction, vol. 106, p. 102895,
2019.

[46] Y. Liu, Q. Wang, H. Hu, and Y. He, "A novel real-time moving target
tracking and path planning system for a quadrotor UAV in unknown
unstructured outdoor scenes," IEEE Transactions on Systems, Man, and
Cybernetics: Systems, vol. 49, no. 11, pp. 2362-2372, 2018.

[47] H. Wu, S. Song, K. You, and C. Wu, "Depth control of model-free AUVs
via reinforcement learning," IEEE Transactions on Systems, Man, and
Cybernetics: Systems, vol. 49, no. 12, pp. 2499-2510, 2018.

[48] Y. Wang, J. Sun, H. He, and C. Sun, "Deterministic policy gradient with
integral compensator for robust quadrotor control," IEEE Transactions
on Systems, Man, and Cybernetics: Systems, pp. 1-13, 2019.

\end{document}